\documentclass[a4paper,12pt]{spieman}  % use this instead for A4 paper
\usepackage{amsmath,amsfonts,amssymb}
\usepackage{graphicx}
\usepackage{setspace}
\usepackage{tocloft}

\title{Calibration unit for the near-infrared spectropolarimeter SPIRou}

\author[a]{Isabelle Boisse}
\author[b]{Sandrine Perruchot}
\author[a, c]{Fran\c{c}ois Bouchy}
\author[b]{Fran\c{c}ois Dolon}
\author[b]{Fran\c{c}ois Moreau}
\author[b]{Rico Sottile}
\author[c]{Fran\c{c}ois Wildi}
\affil[a]{Aix Marseille Universit\'e, CNRS, Laboratoire d'Astrophysique de Marseille UMR 7326, 13388, Marseille, France}
\affil[b]{Observatoire de Haute-Provence, CNRS/OAMP, 04870 Saint-Michel-l'Observatoire, France}
\affil[c]{Observatoire de Gen\`eve,  Universit\'e de Gen\`eve, 51 Chemin des Maillettes, 1290 Sauverny, Switzerland}

\cftpagenumbersoff{figure}
\cftpagenumbersoff{table} 
\begin{document} 
\maketitle

\begin{abstract}
SPIRou is a near-infrared spectropolarimeter and high precision radial velocity instrument, to be
implemented at CFHT in end 2017. It focuses on the search for Earth-like planets around M dwarfs and on
the study of stellar and planetary formation in the presence of stellar magnetic field.
The calibration unit and the radial-velocity reference module are essential to the short- and long-term precision (1~m/s).
We highlight the specificities in the calibration techniques compared to the spectrographs HARPS (at LaSilla, ESO) or
SOPHIE (at OHP, France) due to the near-infrared wavelengths, the CMOS detectors, and the instrument
design.
We also describe the calibration unit architecture, design and production.  
\end{abstract}

% Include a list of up to six keywords after the abstract

% Include email contact information for corresponding author
{\noindent \footnotesize\textbf{*}Isabelle Boisse,  \linkable{isabelle.boisse@lam.fr} \\
\noindent \footnotesize\textbf{*}Sandrine Perruchot,  \linkable{sandrine.perruchot@osupytheas.fr} }

%\begin{spacing}{2}   % use double spacing for rest of manuscript

\section{Introduction}
\label{sect:intro}  % \label{} allows reference to this section
%Rappel SPIRou instrument et objectifs scientifiques \\

% copier du sf2a a r\'{e}crire
%An overview of the key aspects of SPIRou's optical and mechanical
%design is given in \citet{2014SPIE.9147E..15A} and \citet{2013sf2a.conf..497D}.  
%\citet{2013sf2a.conf..509S} detailed several aspects of the science cases and Moutou et al. (2015) presented the SPIRou legacy survey. 

The SPIRou spectropolarimeter is being built in order to become a leading instrument on three forefront science topics, (i) the
quest for habitable Earth-like planets around very- low-mass stars, (ii) the study of low-mass star and
planet formation in the presence of magnetic fields, and iii) the mass determination of transiting planets orbiting low-mass stars. Since M dwarfs are faint in the visible, SPIRou was designed to work in the near-infrared wavelengths (nIR). It is optimised for accurate radial-velocity (RV) measurements on M dwarfs. The polarimetric capability is needed to measure and characterise the structure of magnetic field. The nIR will allow to characterise the topologies of young protostars that are also faint in the visible. Moreover the Zeeman effect increases with
wavelength. A detailed discussions of several aspects of the science cases are given in Delfosse et al. (2013) [1] and Santerne et al. (2013) [2]. Moutou et al. (2015) [3] present the SPIRou legacy survey.

SPIRou should reach a maximum efficiency and optimum precision, providing the heritage derived from HARPS and SOPHIE spectrographs and the Espadons/Narval spectropolarimeters. It will be able to cover a very wide single-shot nIR spectral domain ($0.98-2.35 \mu m$)
at a resolving power of $\sim70,000$. It will produce polarized spectra of low-mass stars with a $15\%$ average throughput and a RV precision of 1~m/s. An overview of the key aspects of SPIRou's optical and mechanical design is given in Artigaud et al. (2014) [4]. SPIRou is being developed by seven countries: Canada, France, Switzerland, Taiwan, Brazil, Portugal, Hawaii and 12 institutes in these countries. The consortium is led by IRAP (Toulouse, France).  SPIRou is made of several sub-systems, each of them being simultaneously developed in the different institutes of the consortium. The first sub-systems of SPIRou are now being delivered to Toulouse, and the integration and tests will take
place at IRAP until mid 2017, when the instrument will be shipped to Hawaii. SPIRou will be mounted on the Canada-France-Hawaii telescope (CFHT) and tested 
at the end of 2017, and open to the CFHT community in 2018. 

Essential to the RV accuracy is the calibration unit and the radial-velocity reference module, located in the Coud\'e room and fiber linked to the spectrograph; this calibration module is being assembled between Geneva and Observatoire de Haute Provence. The calibration unit was just sent to IRAP in May 2016. Other nIR velocimeters are being built in the world, e.g. CARMENES (Quirrenbach et al. 2014 [5]), IRD (Kotani et al. 2014 [6]), HPF (Hearty et al. 2014 [7]). None of these instruments, however, includes the spectropolarimetric capability (required for magnetic field measurements) nor the essential K band (with a large RV content for low mass stars and a large flux for embedded sources). These constraints drove a specific calibration unit for SPIRou.

In this article, after this recall of the science goals and challenges of the instrument, we will detail the constraints given to the calibration unit (Section 2). In section 3, we will expose its technical specifications before to present in Sect. 4 the tests and options that still need to be done and define.

\section{Requirements for the SPIRou calibration unit}

The SPIRou Calibration Unit builds upon the experience of the existing SOPHIE and HARPS spectrographs; class-leaders instruments for high-accuracy RV measurements. While SPIRou shares many characteristics with these instruments, specific challenges are brought by the nIR domain and by the spectropolarimeter design.

\subsection{Heritage from HARPS and SOPHIE}
\label{sect:title}
%description succincte des unit\'{e}s de calibs HARPS et SOPHIE\\

%\subsection{Specifics constraints deriving from the scientific objectives}
%\label{sect:title}
As HARPS and SOPHIE, one of the main scientific objectives of the SPIRou instrument is to search for low-mass planets. A high precision is then requested in the RV measurement. The calibration unit goal is to calibrate and to characterize the spectropolarimeter response to secure the highest possible RV stability, both for short term (one night) and long term (several years) activities. Accurate calibration requires stable and repeatable illumination. Therefore the calibration unit has to provide calibration sources in order to perform the following calibrations:  

  1)  location and geometry of spectral orders,

  2)   the blaze profile and spectral flat-field response (pixels response),

  3)  the slit geometry,

  4)  the wavelength calibration,

  5) the simultaneous drift (zero point). \\
  
\noindent For that, part of the SPIRou calibration unit architecture comes from HARPS@ESO and SOPHIE@OHP heritage:  
\begin{itemize}
%\item For the long term precision, the instrument should have a stabilise point spread function (1), (2). This will be help thanks to an accurate flat field calibration, thanks to the illumination of the spectrograph with a white light source that should have the flattest continuum across the bandwidth. White source allows to determine the Blaze profile and spectral flat-field response
\item The location and geometry of spectral orders (1) and the blaze profile and spectral flat-field response (2) are determined thanks to the illumination of the spectrograph with a white light source that should have the flattest continuum across the bandwidth.
\item The wavelength calibration (4) gives the wavelength value at a pixel position. This relation is determined thanks to the use of a hollow cathode lamp that have a spectrum of unresolved emission lines at known wavelengths. The number of lines should be sufficient per spectral order and cover all the bandwidth. The spectrum should not present thermal background. Fabry-Perot spectra will also be used to locally improve the wavelength solution. %A not too important dynamic of the lines flux is preferred.
\item The spectrograph is in a vacuum chamber controlled in temperature and pressure. However, even a little variation of these two parameters induces a wavelength shift. To get the best accuracy on the RV measurement, the drift of the spectrograph (5) is monitored at the time of the observation. It is done thanks to the simultaneous calibration mode, where the stellar spectra is recorded through the science fibers, at the same time a calibration lamp illuminates the calibration fibers. The flux level of the calibration lamp should be adjusted in order to keep the same level of flux independently of the exposure duration. %The intensity level of the RV reference system should be adjusted as function of the exposure time of the science target in order to keep the same level of flux independently of the exposure duration. 
This is resolved by a density wheel, named "flux balance module" (see Sect. 3.7).
\item In order to measure the instrumental drift, the science target should be observed simultaneously with the RV reference module which is nominally based on a Fabry-Perot etalon or in backup solution to a second Hollow-Cathode lamp. Fabry-perot etalon is being preferred to HC lamps due to the high dynamics of their emission lines.
%To monitor the drift of the spectrograph (5), due to the high dynamic of the emission lines from HC lamps, fabry-perot (RV ref module) is being preferred. In case of deficiency of the RV ref module, a second HC lamp will be used.
%We use two Hollow Cathode Lamps HC1 and HC2. HC1 should be used every day and night for the instrument calibration. HC2 is fixed and will be the long term reference. HC2 should not be changed. in case of deficiency of the RV ref module.

\end{itemize}

\subsection{Constraints from the instrument design}
\label{sect:title}
\begin{figure}
\begin{center}
\begin{tabular}{c}
\includegraphics[height=12cm]{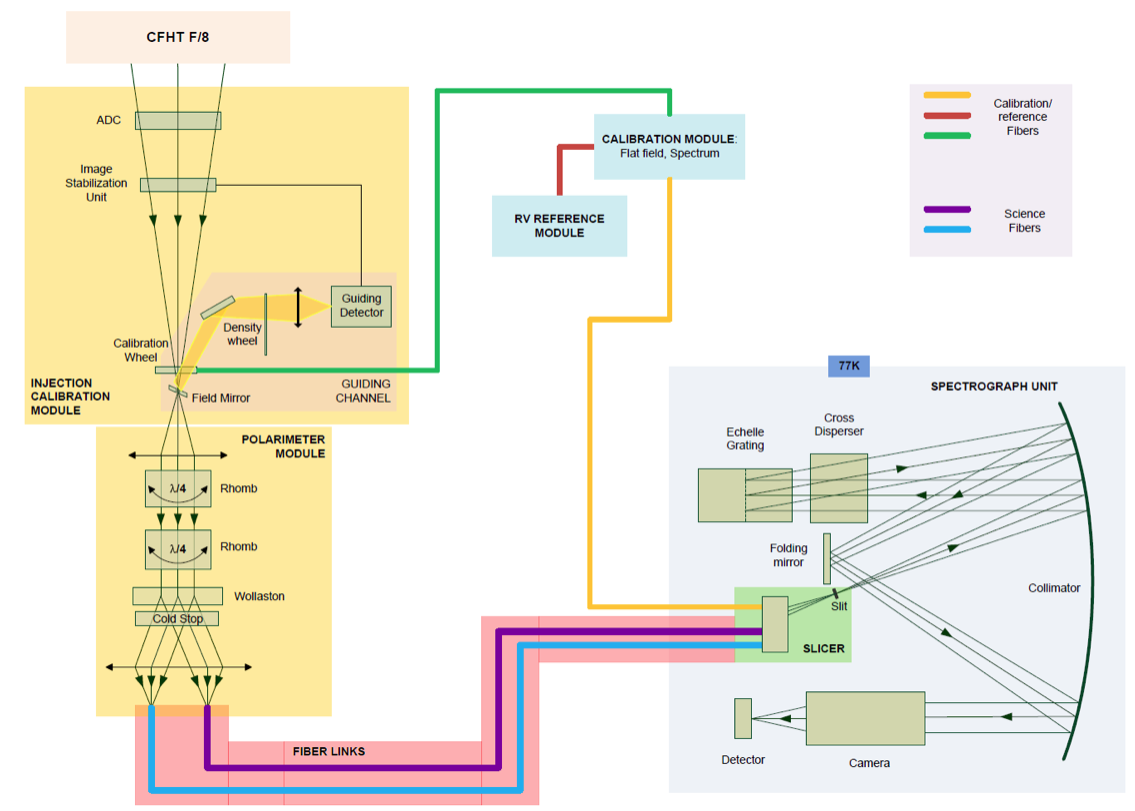}
\end{tabular}
\end{center}
\caption 
{ \label{fig:spirouinstrument}
SPIRou schematic View. } 
\end{figure}

In order to correct from instrumental noise, the calibration light should follow the same path than stellar light. In practice, it means that the calibration light passes through the bonnette of the telescope. But, the polarimeter and two fibers for the two polarizations of the light are already needed in the bonnette. It was then decided to have a specific channel that it goes directly to the spectrograph. The calibration system has then two outputs to feed the spectrograph through two channels (see~Fig.\ref{fig:spirouinstrument}): \\
- The Cassegrain link which uses the same optical path as the target through the Cassegrain Unit, \\
- The Reference link which is used essentially for simultaneous drift measurement, going directly at the spectrograph slicer entrance.\\
 This leads to have two trolleys, each with one fiber. The two links may be illuminated by the calibration system simultaneously (by the same lamp or by different ones) or independently. \\%in order to have the possibility to get two simultaneous calibrations lamps (see Fig.\ref{fig:trolleys}).\\

% It is the reference channel. Another finer is going to the cassegrain bonnette in order to feed the stellar science fibbers with calibration lights. Since we have two different channels, we got two fibers. 
%Calibration light feeds the spectrograph through two channels: one following the stellar beams path through the Cassegrain Unit (Cassegrain fiber link), the other directly at the spectrograph slicer entrance (Reference fiber link). 
%The Cassegrain calibration channel: links directly the calibration module to the Cassegrain unit. The calibration beams follow the same optical path than the stellar beams. The Reference calibration channel: links directly the calibration module to the spectrometer through the slicer unit.

\begin{figure}
\begin{center}
\begin{tabular}{c}
\includegraphics[height=6cm]{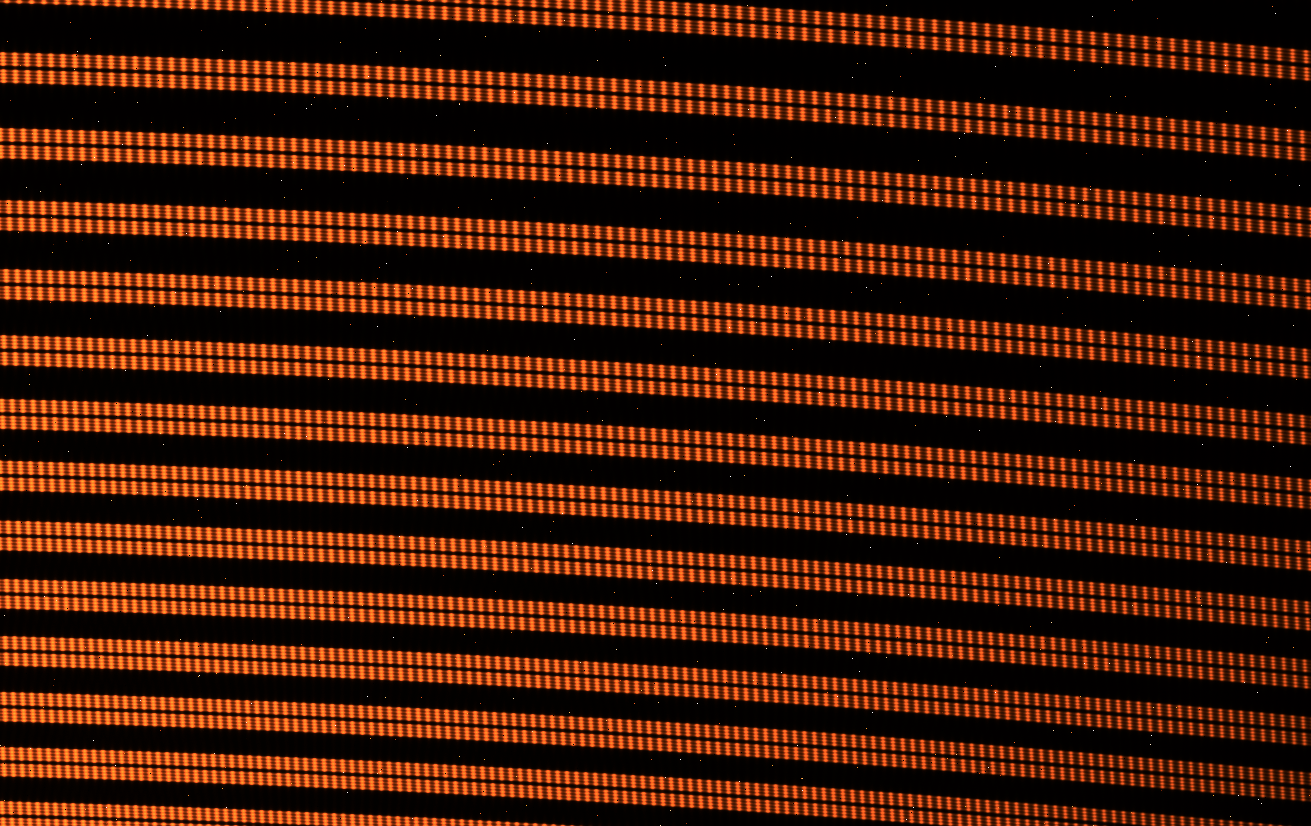}
\end{tabular}
\end{center}
\caption 
{ \label{fig:FP_simu}
Simulated image of the detector when the science fibers are illuminated by the Fabry-Perot etalon. No light is sent through the calibration fiber. Zoom on a part of the detector showing the slit geometry. } 
\end{figure}

\noindent Considering the SPIRou spectrograph R2 grating, the beam coming from each science fiber needs to be sliced and rearranged to a slit to ensure the SPIRou required resolution. The calibration day time would also permit to determine the slit geometry along the orders. This will be done thanks to the analysis of a Fabry-Perot exposure (see Fig. \ref{fig:FP_simu}).

\subsection{Requirements related to the nIR bandwidth}
\label{sect:title}

%\subsubsection{Spectral coverage}
%\label{sect:title}
%- sources (tungstene, amps hollow cathode surement UNe)\\
%- mat\'eriaux transitif ˆ ces longueurs d'onde\\

The calibration lamps shall cover the full wavelength domain of the SPIRou instrument $[0.98 - 2.35 \mu m]$, and all the optics should have a good transmission on all spectral coverage. \\
%\subsubsection{Limiting the thermal background}
%\label{sect:title}

\noindent For all exposures, the emissivity from the instrument shall be reduced at maximum. This is a strong constraint coming from the K-band observation ($2 - 2.35 \mu m$). Specific injection doublet in S-FTM16/CaF2 and fluoride fibers allow to minimize the thermal background along the optical path. 

For science exposure with simultaneous calibration, the thermal background level in the reddest order should not reach the saturation level of the detector. The few last redder spectral orders can be removed for the computation of the spectrograph drift.

For long exposures on faint targets without simultaneous calibration, the Reference channel should not introduce a thermal background level that may pollute the science spectra. The thermal flux from the reference link should be reduced at maximum (goal smaller than 22 ph/s/$\AA$) at spectrograph entrance. For science exposures without simultaneous exposures, the Cold source is selected on the flux balance module, so that the optical path is as short as possible, limiting then the thermal flux generation. It is a cooled diffusing element that can be considered as a cold blackbody, creating a low thermal flux.

\section{Technical description of the calibration unit}
\label{sect:sections}

\subsection{Architecture}
\label{sect:title}
\begin{figure}
\begin{center}
\begin{tabular}{c}
\includegraphics[height=10.1cm]{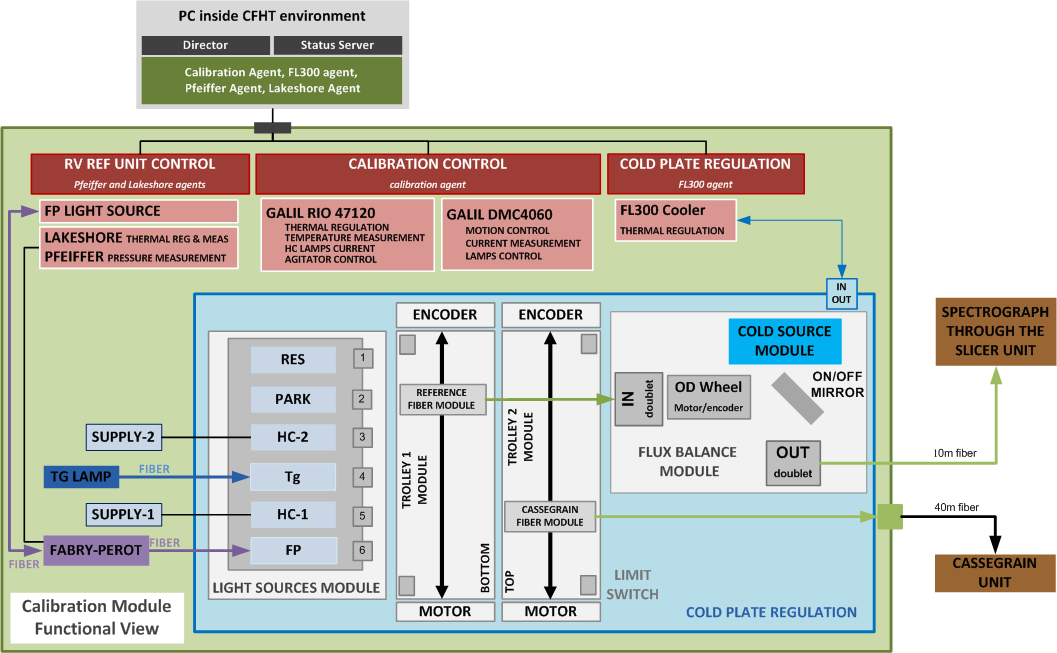}
\end{tabular}
\end{center}
\caption 
{ \label{fig:schemaUC}
Calibration Unit Functional View. } 
\end{figure}

The calibration unit architecture, presented in Fig. \ref{fig:schemaUC}, allows all required functionalities: independent selection of any kind of needed light source for each calibration channel (through Cassegrain unit and science fiber or directly through spectrograph entrance) by trolleys use in front of lamp slots, flux adaptation for simultaneous calibration by circular variable density system on Reference channel, and thermal background limitation on reference channel for long exposures without simultaneous calibration with the help of a cold source module.\\

The calibration module is composed of several optical subsystems:
\begin{itemize}
\item{The calibration lamps, feeding 5 slots: one white lamp, two hollow cathode lamps, the RV reference unit (Fabry-Perot), one reserve.}
\item{The slots, compound of the collimating and filtering optics for each calibration lamp.}
\item{The Reference fiber module and the Cassegrain fiber module (moving parts), imaging the light source plane on each fiber core.}
\item{The flux balance module, only on Reference Channel, to balance flux for simultaneous calibration or select a ÒcoldÓ channel, which is mandatory for long exposures when no calibration light nor thermal background is needed. The selection between a calibration source and the Cold module, is made by the mean of a moveable mirror (ON/OFF mirror). The flux balance is realized with a variable density wheel.}
\item{The two fiber links, conducting light, one directly to the spectrograph (Reference Channel), the other to the Cassegrain Unit where calibration or sky observation can be selected (Cassegrain Channel).}
\end{itemize}

% Note: If compiling with LaTeX+dvipdf, please ensure images generated from 
% other software packages have their bounding boxes set correctly.

\begin{figure}
\begin{center}
\begin{tabular}{c}
\includegraphics[height=9.5cm]{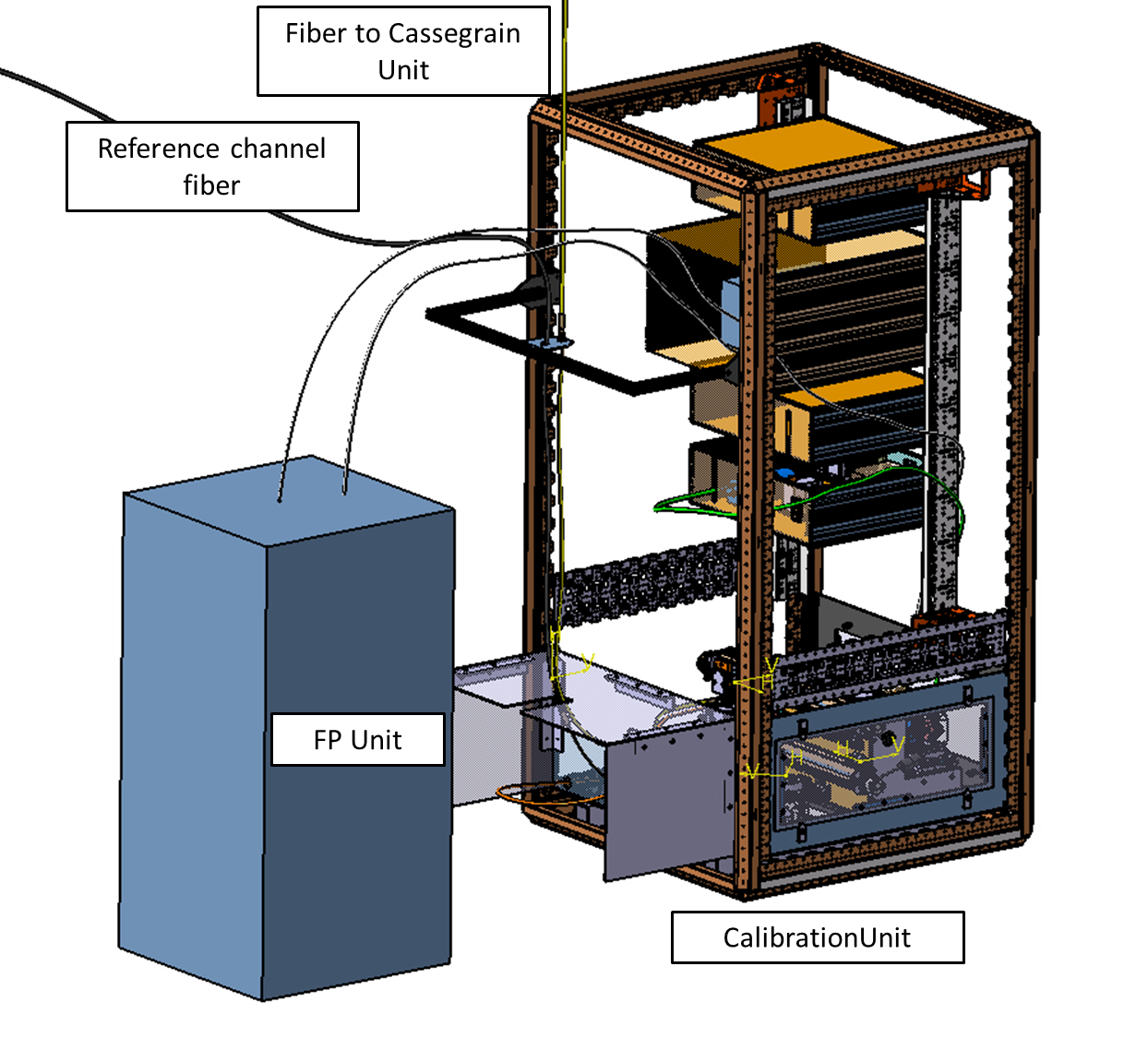}
\end{tabular}
\end{center}
\caption 
{ \label{fig:photoUC}
Calibration Unit schematic View. } 
\end{figure} 

As shown on Fig. \ref{fig:photoUC}, all the calibration Unit stands in a rack except the RV reference unit, described section \ref{sect:FP}, a thermally controlled vacuum tank linked to the rack by a fiber feeding the dedicated slot.

\subsection{Control/command hardware and software}
\label{sect:title}

The calibration Unit is completely remote controlled. The CFHT environment is based on Client-Server concept, where the general client is called "Director" and all the servers are called "agent". The director parse the scripts from the Queued Service Observing (QSO) and send commands to the appropriate agent. All the agents execute the command from the director, drive and control the hardware and send all the status to the Status Server (state of motors, state of lamps, state of regulation, temperatures and currents values for our calibration agents). The control of all the hardware is done through a GALIL controller and RIO Pocket PLC (Programmable Logic Controller).  The Status Server is the data base. It is used to store and display medium term status information with Graphical User Interface.

Spirou calibration agents are developed in accordance with CFHT requirements and constraints.
We distinguish 4 agents for the calibration unit :
\begin{itemize}
\item{Calibration agent ensures the whole control command of the calibration,}
\item{FL300 (ColdPlate Cooler) agent ensures the cold plate thermal regulation of the calibration,}
\item{Pfeiffer agent ensures the monitoring of the RV reference module pressure,}
\item{Lakeshore agent ensures the monitoring of the RV reference module temperature.}
\end{itemize}

\begin{figure}
\begin{center}
\begin{tabular}{c}
\includegraphics[height=10cm]{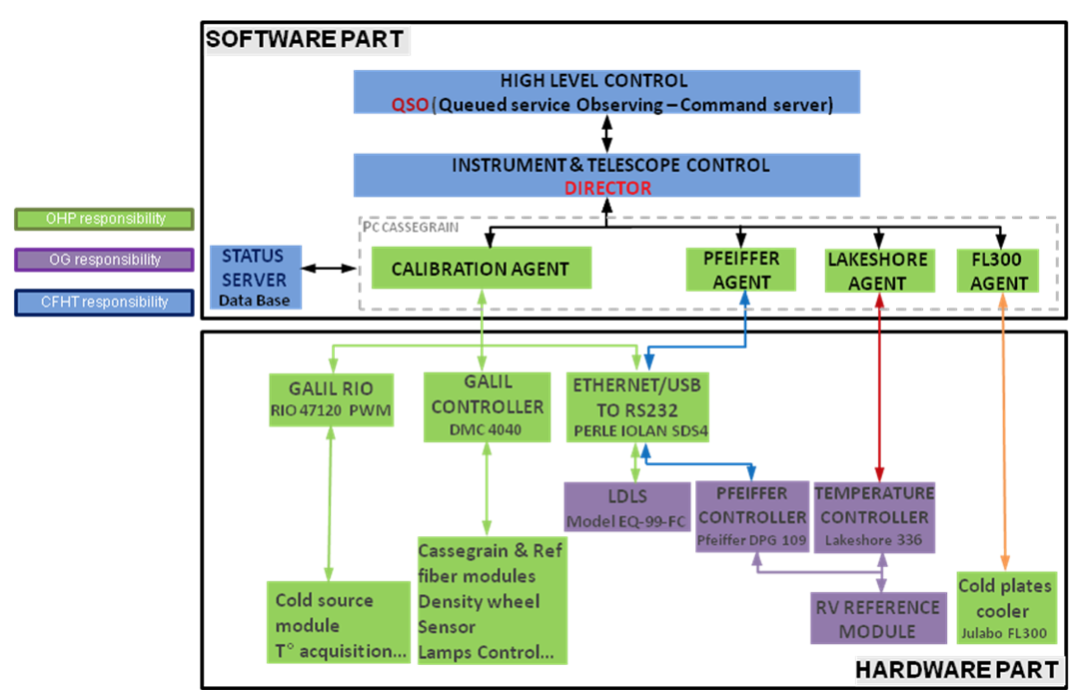}
\end{tabular}
\end{center}
\caption 
{ \label{fig:spirousoftware}
Control/command hardware and software for the Calibration Unit. } 
\end{figure}

\subsection{Fabry-Perot - Radial-velocity reference module}
\label{sect:FP}

SPIROU, just like the HARPS [8],[9], SOPHIE and CORALIE spectrographs use the so-called simultaneous calibration method to reach maximal RV performance: the spectrograph has a doubled fiber input and produces two spectra on the detector: fiber A for the stellar spectrum (calibrated in wavelength by the calibrator) and fiber B for the reference spectrum which is measured during the calibration phase and also during the science exposure in order to track and correct the spectrograph drifts.
The RV reference spectrum has to be highly stable. It must cover the full wavelength range of the spectrograph; its lines are not resolved at the spectrograph resolution so that the spectrograph is insensitive to evolution of the lines profiles. There must be as many lines as possible in the spectral range to reduce the localization noise associated with each line. Of course, for stability, the environment has to be mechanically and thermally very stable and the index of refraction in the gap constant [10]. 

\begin{figure}
\begin{center}
\begin{tabular}{c}
\includegraphics[height=8cm]{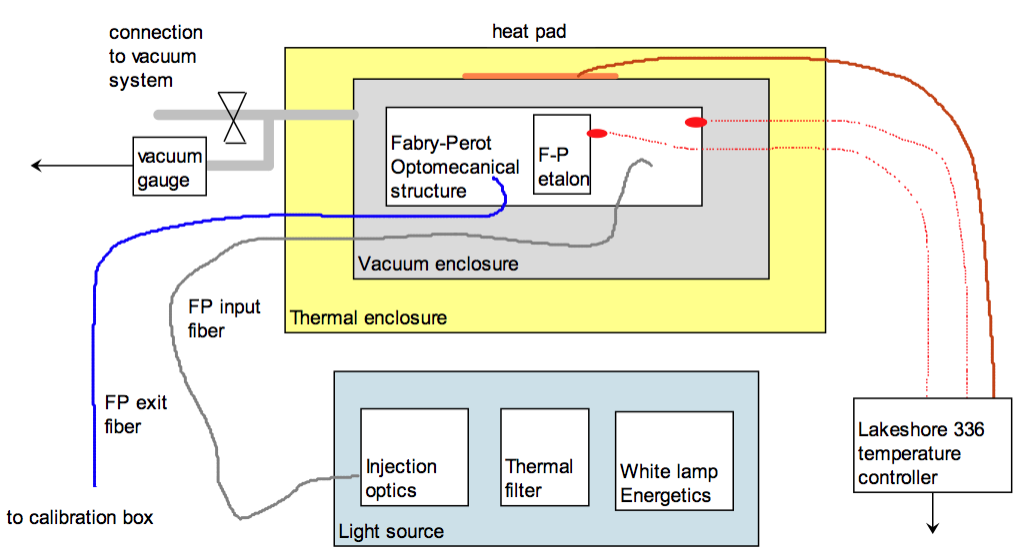}
\end{tabular}
\end{center}
\caption 
{ \label{fig:fparchitecture}
Functional diagram of the complete Fabry-Perot calibrator system. } 
\end{figure}

\noindent The RV reference unit is made of the following elements (see Fig.\ref{fig:fparchitecture}):
\begin{itemize}
\item{As a primary light source we are now using a Laser Driven Light Source (LDLS) which is in essence a Xe lamp excited by a pump laser. It provides very high radiance broadband light closely corresponding to a 10Õ000K blackbody. This lamp is turned on at all times for stability reasons.}
\item{A Fabry Perot (FP) etalon. Etalon means fixed gap. }
\item{Reflective collimator and fiber injection lens.}
\item{$200 \mu m$ diameter optical fiber for light injection into the FP system.}
\item{$600 \mu m$ diameter optical fiber connecting the FP assembly with the rest of the calibration unit.}
\item{A vacuum enclosure holding the FP assembly. An operating pressure below $10^{-3}$mBar is required to insure $10^{-10}$ RV stability. (This is because the index of refraction of air will modify the free spectral range of the etalon).}
\item{A temperature controller to stabilize the FP assembly temperature.}
\end{itemize}

\begin{figure}
\begin{center}
\begin{tabular}{c}
\includegraphics[height=4.8cm]{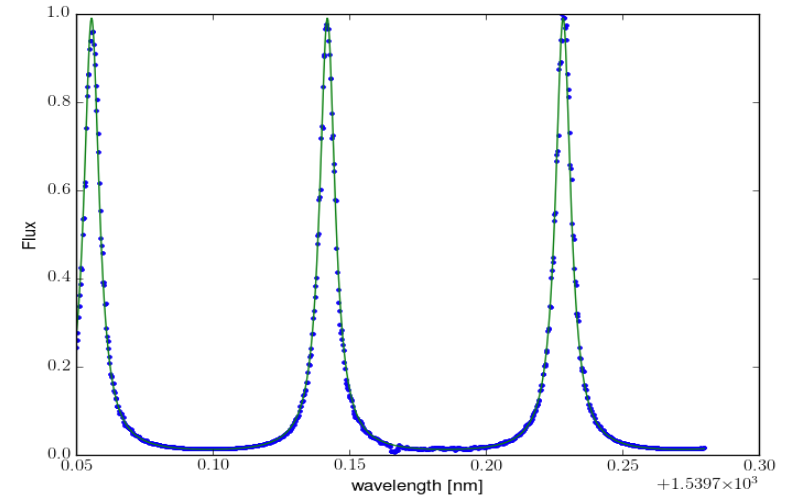}
\includegraphics[height=5.2cm]{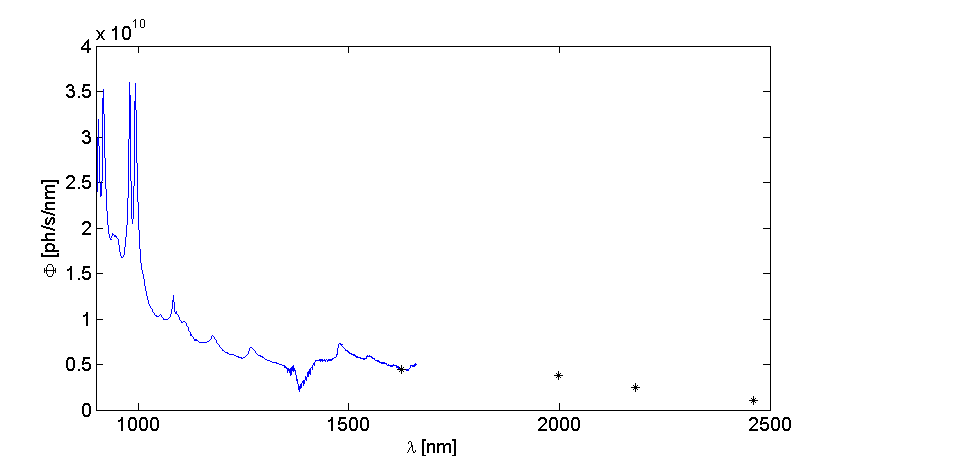}
\end{tabular}
\end{center}
\caption 
{ \label{fig:fpfinesse}
\textit{Left}: Scan of the RV reference module transmission function T($\lambda$). The fitted finesse is 12.7. \textit{Right}: Flux of the RV reference module over the Spirou range. } 
\end{figure}

Regarding the primary lamps, two solutions where evaluated: A regular and inexpensive quartz tubgsten halogen lamp (QTH) and a laser driven light source (LDLS). The 1st one has a smooth spectrum but a low radiance and the 2nd one has a radiance at least an order of magnitude higher at the expense of a spectrum featuring relatively sharp lines around the pump laser wavelength (see Fig. \ref{fig:fpfinesse}).

One has to be careful when injecting the calibration light into the system. Due to the imperfect scrambling of the multimode fibers, variations in the injection from the calibrator into the fibers going to the spectrograph will induce a variation of the illumination pattern on the fiber (See [11] and [12]). It is highly desirable to have a source with a large etendue to overfill the fiber and be insensitive to alignment variations and that make supercontinuum lasers unsuitable for the task.

Figure \ref{fig:fpfinesse} shows a small part of the transmission spectrum of the RV reference module recorded by scanning a single frequency laser in temperature. The fitted finesse is 12.7. The average spectral flux produced by the RV reference was also measured. From 950 to 1700nm, this was done with a medium resolution spectro-photometer. Beyond the range of this instrument, we have used discreet bandpass filters hence the 4 points in the figure below at $\lambda$=1625.7~nm, $\lambda$=1999.5~nm, $\lambda$=2180.2~nm and $\lambda$=2460.8~nm. One can note four sharp spike at 900 and 1000nm. Test at system level will determine if this is tolerable or if it is better to revert to the QTH source.

\subsection{Slots and trolleys}
\label{sect:title}

\begin{figure}
\begin{center}
\begin{tabular}{c}
\includegraphics[height=9cm]{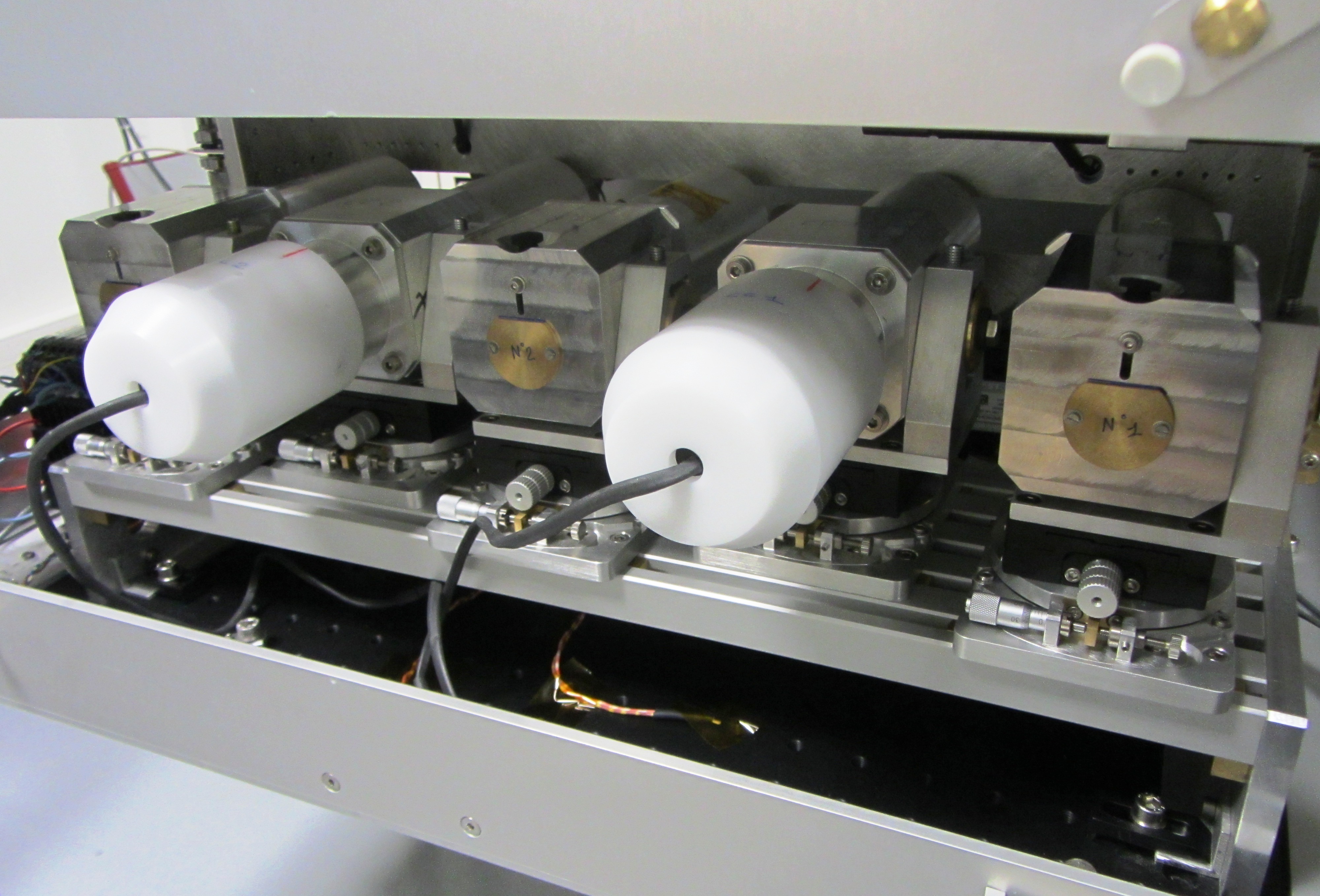}\\
\includegraphics[height=11.38cm]{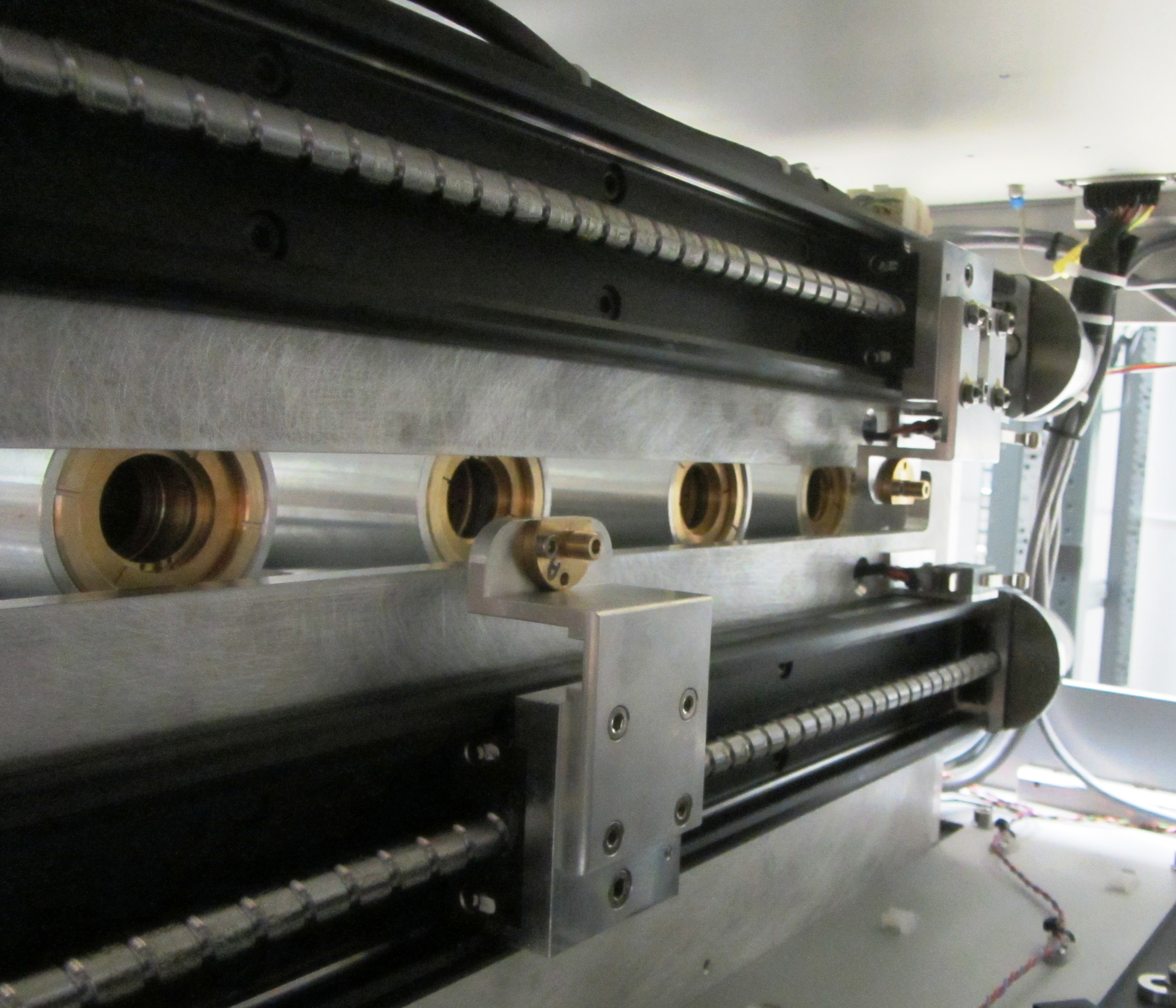}
\end{tabular}
\end{center}
\caption 
{ \label{fig:trolleys}
Views of the slots (\textit{top}) and trolleys (\textit{bottom}) (fibers not connected). } 
\end{figure}

The slot system is dedicated to inject fibers with calibration sources light (continuum, Hollow cathodes, RV reference) at working aperture. It images the source plane (a diffuser for continuum and RV reference module, the cathode for the Hollow cathodes), on the fiber cores with a 1:10 relay composed by lenses with focal lengths 125 and 12.5 mm and a single layer coating. The two moving fibers are independently mounted on each trolley at the focus of their own 12.5 mm focal length - focusing doublet, the optical axes of the two systems being separated by 12.7 mm in front of a lamp slot (see Fig.\ref{fig:trolleys}). Image quality is not a demanding requirement for calibration, since itÕs about inject spectral flux in fibers, not images. Then commercial optics, even not optimized for NIR, are chosen for slots, with adequate coating for the domain (MgF2 single layer). Each source plane (diffuser, cathode) imaged on each fiber, beam collimated out of each slot.

The trolleys include two actuators (linear modules, sensors, motors, encoders, motion controller) to independently position with precision, accuracy and repeatability the fiber injection systems in front of the selected calibration lamp. The moving fibers are guided in the cumbersome module to preserve their integrity during instrument lifetime, forbidding misrouting or exaggerated curvature radius.

\subsection{Fibers}
\label{sect:title}

\begin{figure}
\begin{center}
\begin{tabular}{c}
\includegraphics[height=9cm]{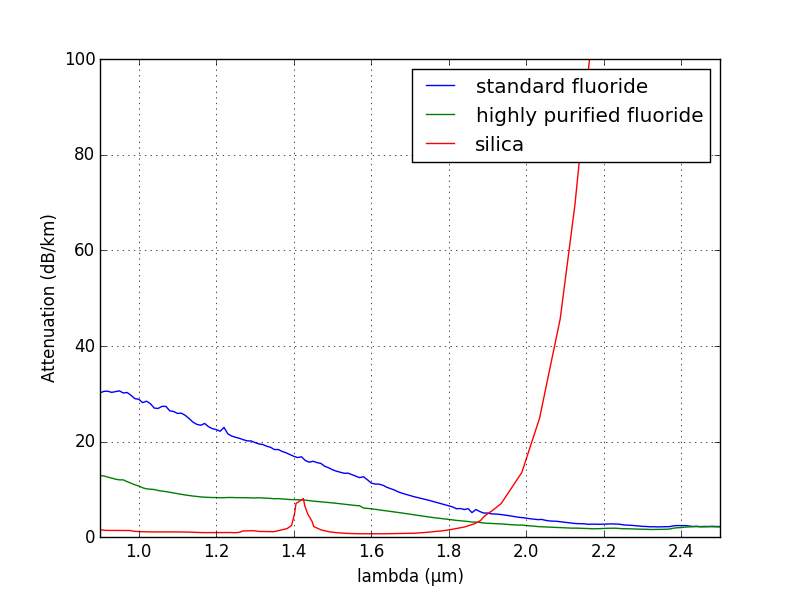}
\end{tabular}
\end{center}
\caption 
{ \label{fig:fibers}
Attenuation for typical silica fiber and fluoride fibers (from Le Verre Fluor\'e). } 
\end{figure}

The SPIROU spectral domain [$0.98\mu$m Ð $2.35\mu$m] leads to some changes from HARPS or SOPHIE context, considering material transmittance but also emissivity. For calibration, flux level is not so much an issue compared to spectral flux balance for SNR considerations. This last constraint is especially an issue for the fiber material. Therefore fluoride glass is mostly preferred to silica because of the very high attenuation of the silica in the K-band ($\sim1000$ dB/km at $2.35\mu$\,m), which is unacceptable. The longest fibers (40 m and 10 m) benefit from the purification effort made for the science fibers of the project by Le Verre Fluor\'e (see Fig.\ref{fig:fibers}).

The Numerical Aperture (NA) of the fibers is limited to 0.15 when manufactured to limit the thermal background generated outside the useful aperture that could diffuse through the system. The core shapes are simply circular, the scrambling devices necessary to assure sufficient long term stability being located just at the spectrograph entrance for both channels. Cables are from different natures considering protection when exposed on the route, or weight when moving.

\subsection{Cold source module}
\label{sect:title}

The Reference channel permanently illuminates the spectrograph, even when no calibration is demanded. For long exposures on faint targets without simultaneous calibration, this channel should not introduce a thermal background level that may pollute the science spectra. Then the thermal flux from the reference link should be smaller than around 30 ph/s/$\AA$ at spectrograph entrance. %At $2.35\mu m$ ?

\begin{figure}
\begin{center}
\begin{tabular}{c}
\includegraphics[height=6.5cm]{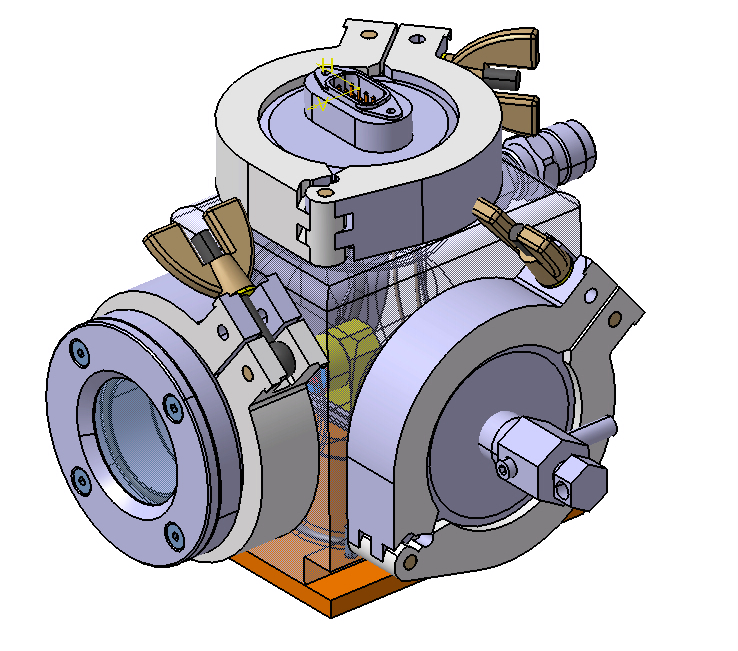}
\hspace{0.7cm}
\includegraphics[height=5.9cm]{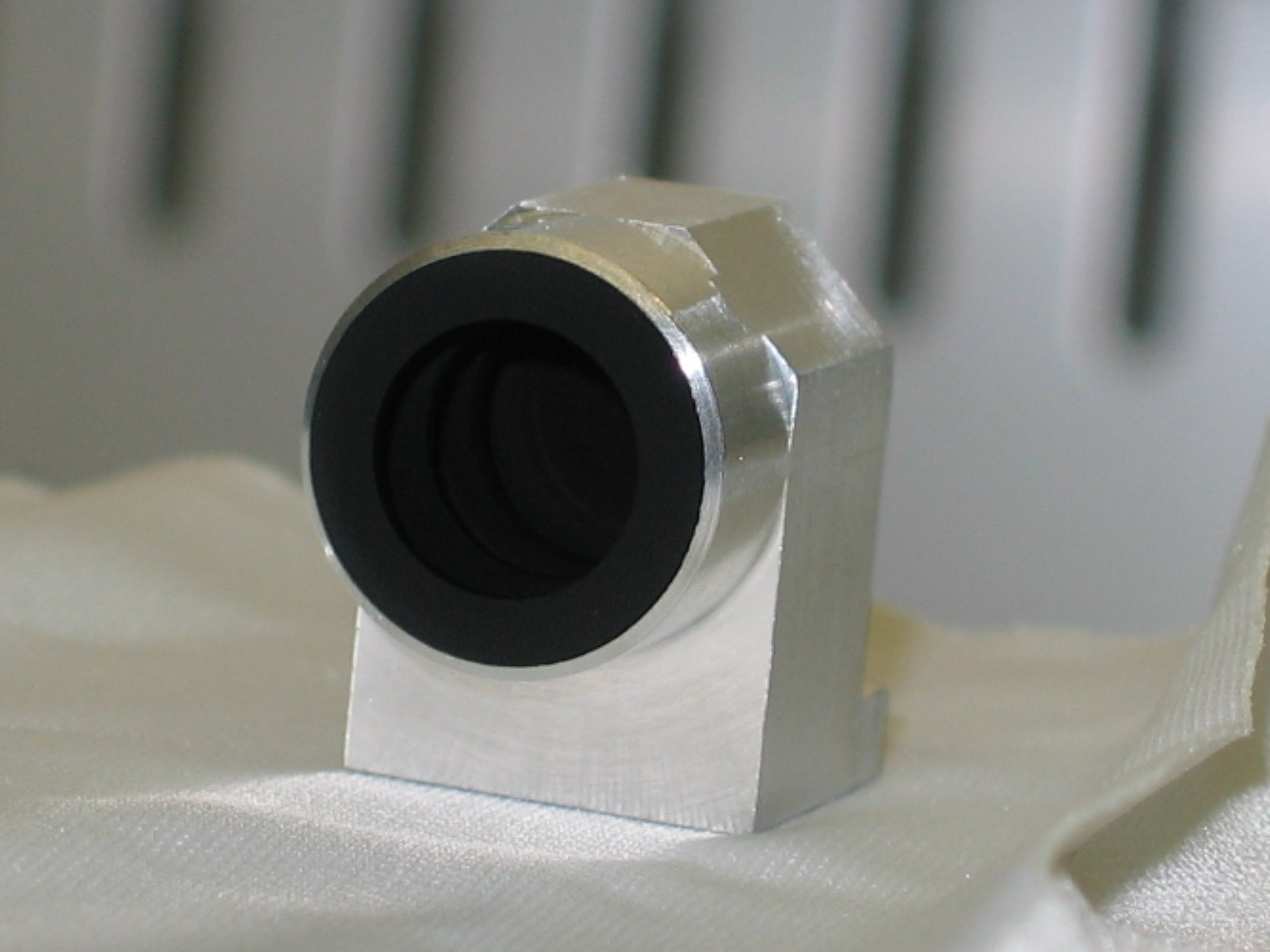}
\end{tabular}
\end{center}
\caption 
{ \label{fig:cold}
\textit{Left:}  Schematic view of the cold source module. 
\textit{Right:} View of the cold part.} 
\end{figure}

A dedicated channel has been designed to reach such level, minimizing the number of contributing elements. A flip mirror allows to choose between this cold channel or calibration, the mirror being off beam in Cold case. Two concepts were possible: using the Narcissus effect with the SPIRou detector or creating a cold source. This last concept has been chosen. The cold source module is composed of a surface with a high emissivity (Black Acktar, see Fig. \ref{fig:cold}) placed in a cell at low temperature closed by a very transmissive especially coated window in thermal wavelength range. The cell is filled with neutral gas and cooled to $-25^{\circ}$C by a thermoelectric module. The module is in contact with a thermally controlled breadboard at $+10^{\circ}$C to evacuate heat. The cold target is viewed by the Reference fiber and then the spectrograph through a specific CaF2/S-FTM16 doublet to minimize the absorbance factor. This last element is also coated to minimize infrared losses and then also the possible straylight. As it is mounted on the thermally controlled breadboard, it benefits from its low temperature to reduce its residual thermal emission.

 It is also necessary to keep optics cold, typically at $13^{\circ}$C thanks to the cold plate for very long exposures. In order to keep the temperature very stable and homogeneous inside the cold plate and the LSU, it is recommended to never switch off the regulation.

As mentioned before, the fluoride fiber is of high quality and designed with a reduced NA of 0.15 to reduce its thermal background (see [13]).

\subsection{Flux balance module}
\label{sect:title}

In order to measure the instrumental drift, the science target should be observed simultaneously with the RV reference module which is nominally based on a Fabry-Perot etalon or in backup solution the Hollow-Cathode lamp. The intensity level of the RV reference system should be adjusted as function of the exposure time of the science target in order to keep the same level of flux independently of the exposure duration. Furthermore, the thermal background level in the reddest order (K band) should not reach the saturation level of the detector. The few last redder spectral orders can be removed for the computation of the spectrograph drift.

This is done by a continuous variable density wheel placed on a collimated beam generated by a relay, which images the fiber output from the slots onto the reference fiber linked to the spectrograph. The 1:1 relay is a symmetric system made of two specific doublets, the output one being also used for Cold Channel described in previous section. The whole system is mounted on the cooled breadboard preventing an excessive thermal background emission. The density system is made of two identical counter-rotating continuous variable densities to keep the far-field homogeneous.

\section{Fine-tuning tests and monitoring of the calibration unit performances}
%- d\'{e}finition de la s\'{e}quence de calibs de jours\\
%- choix de la lampe HC\\
%- les temps de pose, \\
%- le courant des lampes HC\\

%- link with the DRS, catalogue de raies\\
%The data reduction pipeline is being developed by the consortium (effort led by LAM, Marseille) and will
%be delivered to CFHT with the instrument, aiming at a complete extraction of spectra as well as radial-velocity
%and polarisation signatures available a few minutes after the observations.

Several choices, needed to get the best accuracy on the stellar spectrum and its derived RV and magnetic properties, could be done only with the complete instrument: spectrograph resolution, science detector, data reduction system (DRS) and other parts between the calibration unit and the SPIRou detector. One of the main reasons is that SPIRou is forefront in the exploitation of the K band and the detector H4RG will be the first of its type and its performances are not well known at that moment. The DRS pipeline is being developed by the consortium (effort led by LAM, Marseille) and will be delivered to CFHT with the instrument, aiming at a complete extraction of spectra as well as RV and polarisation signatures available a few minutes after the observations.\\

\noindent Hence, several tests will be done at IRAP and CFHT:
 \begin{itemize}
\item Final choice of the Hollow cathode types after test to determine the effective number and repartition of usable lines. The hollow cathodes lines catalog is mandatory to determine the accuracy of the wavelength calibration solution. The definition of the catalog is an iterative process, needing some spectral images with the definitive spectrograph system.  
\item Select the warming time for the HC lamps. To reach the highest spectral stability, Hollow-Cathode lamp should warm and should be turn on few minutes before being used for calibration as it is done on other spectrographs like HARPS or SOPHIE. 
\item Select the Hollow cathode lamps supply current optimization, depending on each lamp, regarding metallic/gas flux ratio and flux stability (see [14]).
\item Select the optimal exposure time for all the sources (with or without neutral density for HC lamps). Determination of the optimal exposure time with HC source with added ND. The exposure time for the calibration should be negligible with respect to the instrumental drift time-scale (few hours) but not too short with respect to possible high frequency vibrations (few Hz). Therefore, calibration exposure time should be in between 5 and 100 s nominally and in any case shorter than 1800 s. 
\item Check diffused light or ghost on the spectrograph due to strong emission lines in and out-side spectral range.
\item Check the thermal background on the spectrograph detector with maximal exposure time with HC source (No saturation in red orders for simultaneous reference)  
\item Select the white lamp for the RV reference module. 
%\item Flux level adjustment for each of the 4 sources, through exposure times optimization and neutral densities addition in internal calibration box light paths, OD being determined lately (but as soon as possible, other instrument parts needed).\\
\end{itemize}

 % the choice of the HC lamp will depend on its comportment in the K-band. We conducted several tests on visible spectrographs, gathered thanks to the PI team spectrum on GIANO (ref) instrument and thanks to lower resolution spectrograph,... but GIANO ˆ la band K...

When the instrument will be on the sky:
\begin{itemize}
\item The DRS will monitor the flux of the HC lamps that have a lifetime of around 10,000~hours. The flux increase with the ageing of the lamp. This increase could be reduce thanks to a lower supply current, but this will modify the flux ratio between the gas and the metallic lines.
\item The calibrations for the determination of location and geometry of spectral orders, the slit geometry, the blaze profile, the spectral flat-field response and the wavelength calibration should be performed during the day to meet the requirement of observation availability time during the night. This is possible with the assumption of an instrument drift smaller than 1~m/s during a whole night. %The daily calibrations should take less than 2 hours and should be completed at least 2 hours before the start of the night to avoid any persistent effect on the IR detector.   
 The daily calibrations should take less than 2 hours and should be completed at least 2 hours before the start of the night. This is to prevent permanence effect on the detector (particular sensitivity due to the CMOS detector). This is another reason to prefer Fabry-Perot etalon to HC lamp for simultaneous calibration during the night. Emission lines of HC have important dynamics with strong gas line that we anticipate that some of them will saturate the detector.
\item The complete calibration sequence will also be done at the end of the night. 
\end{itemize}

%SPIRou is a near-infrared spectropolarimeter, designed for the detection of exoplanet around low-mass stars, and for the detection of magnetic fields of young stellar objects. It is currently under construction, and should give its first light in fall 2017, in order to be available to the community in 2018.  

\acknowledgments 
The authors thanks the financial support by the Laboratoire d'Astrophysique de Marseille and the OSU Pyth\'eas. This work has been carried out thanks to the support of the A*MIDEX project (n$^{\circ}$ANR-11-IDEX-0001-02) funded by the "Investissements d'Avenir" French Government program, managed by the French National Research Agency (ANR). F. Wildi gratefully acknowledges the outstanding support of Bruno Chazelas and Federica Cersullo of the University of Geneva in the integration and test of the Radial Velocity Reference module. \\

%%%%% References %%%%%
\noindent \textit{References}

\noindent 
$[1]$ Delfosse X., Donati, J.-F., Kouach, D. et al., " World-leading science with SPIRou - The nIR spectropolarimeter / high-precision velocimeter for CFHT", SF2A, 497 (2013)\\
$l2]$ Santerne, A., Donati, J.-F., Doyon, R. et al., " Characterizing small planets transiting small stars with SPIRou", SF2A, 509 (2013)\\
$[3]$ Moutou, C., Boisse, I., HŽbrard, G. et al, " SPIRou: a spectropolarimeter for the CFHT", SF2A, 205 (2015)\\
$[4]$ Artigaud, E., Kouach, D., Donati, J.-F. et al., " SPIRou: the near-infrared spectropolarimeter/high-precision velocimeter for the Canada-France-Hawaii telescope", SPIE, 9147, 15 (2014)\\
$l5]$ Quirrenbach, A., Amado, P.J., Caballero, J.A. et al. "CARMENES instrument overview", SPIE, 9147, 1 (2014)\\
$[6]$ Kotani, T., Tamura, M., Suto, H. et al., " Infrared Doppler instrument (IRD) for the Subaru telescope to search for Earth-like planets around nearby M-dwarfs", SPIE, 9147, 14 (2014)\\
$[7]$ Hearty, F., Levi, E., Nelson, M. et al., " Environmental control system for Habitable-zone Planet Finder (HPF)", SPIE, 9147, 52 (2014)\\
$[8]$ Cosentino, R., Lovis, C., Pepe, F. et al, ÒHarps-N: the new planet hunter at TNGÓ,. in Ground-based and Airborne Instrumentation for Astronomy IV , SPIE [8446-66] (2012)\\
$[9]$ Mayor, M., Pepe, F., Queloz, D. et al. ÒSetting new standards with HARPSÓ, The Messenger, 114, 20 (2003)\\
$[10]$ Wildi, F., Pepe, F., Chazelas, B., Lo Curto, G. "A Fabry-Perot calibrator of the HARPS radial velocity spectrograph: performance report "  in Ground-based and Airborne Instrumentation for Astronomy III , SPIE 7735-181 (2010)\\
$[11]$ Chazelas, B., Pepe, F., Wildi, F. Ç Optical fibers for precise radial velocities: an updateÓ, SPIE [8450-124] (2012)\\
$[12]$ Chazelas, B., Pepe, F., Wildi, F., Bouchy, F. ÒStudy of optical fibers scrambling to improve radial velocity measurementsÓ, Modern Technologies in Space- and Ground-based Telescopes and Instrumentation, SPIE 7739-191 (2010)\\
$[13]$ A.Zur and A.Katzir, "Theory of fiber optic radiometry, emissivity of fibers,and distributed thermal sensors", Appl. Opt. 30, 660-673 (1991)\\
$[14]$ Sarmiento, L. F., Reiners, A., Seemann, U et al. "Characterizing U-Ne hollow cathode lamps at near-IR wavelengths for the CARMENES survey", SPIE, 9147, 54 (2014)\\

%\bibliography{report}   % bibliography data in report.bib
%%bibliographystyle{spiejour}   % makes bibtex use spiejour.bst

%%%%% Biographies of authors %%%%%

%\vspace{2ex}\noindent\textbf{First Author} is an assistant professor at the University of Optical Engineering. He received his BS and MS degrees in physics from the University of Optics in 1985 and 1987, respectively, and his PhD degree in optics from the Institute of Technology in 1991.  He is the author of more than 50 journal papers and has written three book chapters. His current research interests include optical interconnects, holography, and optoelectronic systems. He is a member of SPIE.

%\vspace{1ex}
%\noindent Biographies and photographs of the other authors are not available.

%\end{spacing}
\end{document}